\documentclass{aa}  

\usepackage[varg]{txfonts}
\usepackage{graphicx}
\usepackage{orcidlink}
\usepackage{subfig}
\newcommand{\co}{\mbox{\rm $^{12}$CO}}
\newcommand{\coa}{\mbox{\rm $^{13}$CO}}
\newcommand{\cob}{\mbox{\rm C$^{18}$O}}

\newcommand{\kms}{\mbox{\rm km s$^{-1}$}}
\newcommand{\htwo}{\mbox{\rm H$_2$}}

\newcommand{\cmsq}{\mbox{\rm cm$^{-2}$}}
\newcommand{\msun}{\mbox{\rm M$_\odot$}}

\newcommand{\vlsr}{\mbox{V$_{{\rm LSR}}$}}
\newcommand{\VA}{\mbox{V$_{{\rm A}}$}}

\newcommand{\ZCO}{\mbox{V$_{\rm 12}$(v)}}
\newcommand{\ZCOA}{\mbox{V$_{\rm 13}$(v)}}
\newcommand{\PRS}{\mbox{\it PRS}}

\begin{document} 

\title{The SIMPLIFI Survey: Polarization of Dense Filaments with SOFIA}
\subtitle{
III. Relative orientations of antenna temperature gradients with the magnetic field in the DR21 molecular cloud}

\author{
\orcidlink{0000-0002-3871-010X}Mark Heyer\inst{1}
\and
\orcidlink{0000-0003-2133-4862}Thushara G.S.  Pillai\inst{2}
\and 
\orcidlink{0000-0002-5094-6393}Jens Kauffmann\inst{2}
\and
\orcidlink{0000-0002-0294-4465}Juan D. Soler\inst{3}
\and
\orcidlink{0000-0002-0368-9160}Daniel Seifried\inst{4}
\and 
\orcidlink{0000-0002-4666-609X}Laura M. Fissel\inst{5}
\and 
\orcidlink{0000-0003-2020-2649}Gabriel A.P. Franco\inst{6}
\and 
\orcidlink{0000-0002-0368-9160}Paul F. Goldsmith\inst{7}
\and 
\orcidlink{0000-0002-2885-1806}Philip C. Myers\inst{8}
\and 
\orcidlink{0000-0002-5135-8657}Helmut Wiesemeyer\inst{9}
\and
Karl M. Menten\thanks{Deceased}\inst{9}
}

\institute{Department of Astronomy, University of Massachusetts, Amherst, MA 01003, USA \\
\email{heyer@astro.umass.edu} 
\and 
Haystack Observatory, Massachusetts Institute of Technology, 99 Millstone Road, Westford, MA, 01886, USA 
\and
University of Vienna, Department of Astrophysics, T\"urkenschanzstrasse 17, 1180 Wien, Austria
\and 
Universit\"at zu K\"oln, I. Physikalisches Institut, Z\"ulpicher Str. 77, D-40937 K\"oln, Germany
\and
Department of Physics, Engineering Physics and Astronomy, Queen's University, Kingston, ON K7L 3N6, Canada
\and
Departamento de F\'isica, ICEx, UFMG, Av. Ant\^onio Carlos 6627, 30.270-901 
Belo Horizonte, Brazil
\and
Jet Propulsion Laboratory, California Institute of Technology, 4800 Oak Drive Pasadena, CA 91109, USA
\and
Center for Astrophysics Harvard and Smithsonian, 60 Garden St, Cambridge, MA 02138, USA
\and
Max-Planck-Institut f\"ur Radioastronomie, Auf dem H\"ugel 69, D-53121, Bonn, Germany
}
\date{Received XXX; accepted YYY}
\titlerunning{Relative Orientations in the DR21 molecular cloud}
\authorrunning{M. Heyer et al.}

 
\abstract{We investigate magnetohydrodynamic (MHD) turbulence in the DR21 molecular cloud 
using far infrared polarimetry from {\it SOFIA HAWC+} and molecular spectroscopy from the Nobeyama 45~m telescope.
The Velocity Channel Gradient (VChG) method is applied to thin velocity slices of the 
\co\ and \coa\ J=1-0 emission to determine the antenna temperature gradients and their orientations
in the plane of the sky. The relative orientations of the gradients with respect to the magnetic field position angles 
enable an evaluation of theoretically predicted alignment of 
elongated, turbulent eddies for Alfv\'enic, fast and slow modes and self-gravity. 
The projected Rayleigh statistic is calculated  for both the 
full cloud and 6 regions of interest to provide a quantitative assessment of these alignments. 
Our analysis finds mostly parallel alignment
of the gradient direction and the local magnetic field orientation in the Main Ridge of DR21
that comprises most of the cloud mass.  While parallel alignment can result from fast MHD modes or 
cloud collapse, previous studies show spectroscopic evidence for the large scale collapse of the 
DR21 cloud, which supports self-gravity as the primary cause of the observed parallel alignment. 
One of the six 
sub-filaments that are connected to the Main Ridge exhibits  mostly 
perpendicular alignment of the position angles of the gradient and magnetic field. 
These perpendicular relative orientations may be a product of 
Alfv\'en and slow modes in a magnetized, sub-Alfv\'enic  medium even if the sub-filament is 
accreting onto the Main Ridge. 
}


\maketitle
\nolinenumbers

\section{Introduction}
The dynamics of the molecular interstellar medium in the Milky Way 
are characterized by fluctuating densities and velocities over spatial scales ranging from $\sim$0.001~pc
 to tens of pc. 
For scales larger than 1~pc, the Mach numbers are generally greater than 2, indicative of supersonic turbulence. 
But such gas motions also interact with the interstellar 
magnetic field that threads molecular clouds.   This interaction can modify the characteristics 
of the flow if the motions are sub-Alfv\'enic, $\sigma_v/$\VA $<$1, where $\sigma_v$ is the velocity dispersion and 
\VA=|{\rm B}|/$(4\pi\rho)^{1/2}$ is the Alfv\'en 
velocity, $|{\rm B}|$ is the magnetic field strength, and $\rho$ is the mass volume density of the gas.   Under sub-Alfv\'enic
conditions, the turbulent eddies are predicted to become elongated and aligned with the orientation of the magnetic field \citep{Goldreich:1995, Lazarian:1999, Hu:2020}. 
Observationally, this feature of magnetohydrodynamic turbulence is recognized as an anisotropy of 
velocity and density structure functions and the alignment of localized gradients of gas properties with 
the local magnetic field orientation \citep{Heyer:2008, Esquivel:2011, Heyer:2012,Burkhart:2014}.  Such gas properties include 
molecular column densities \citep{Planck:2016}, velocity centroids \citep{Gonzalez:2017}, and molecular line brightness 
temperatures over narrow velocity channels \citep{Lazarian:2018, Hu:2019, Heyer:2020}.  

The DR21 molecular cloud has been a popular and important target for studies of the molecular interstellar medium and star formation.  It is located 
within the Cygnus star forming complex \citep{Downes:1966, Dickel:1978} at a heliocentric 
distance of 1.4$\pm$0.08~kpc \citep{Rygl:2012}.
DR21 is superposed upon the W75 molecular cloud, also residing within the Cygnus complex, that may 
indicate the two clouds are colliding \citep{Dickel:1978}.   As measured by the {\it Herschel Observatory} 
using 
PACS and SPIRE bands, the total mass of the cloud is 2$\times$10$^4$ \msun\ \citep{Hennemann:2012}.
The massive clump, DR21 Main,
 harbors a compact HII region. 
This star formation activity powers a massive (3000~\msun), energetic (2$\times$10$^{48}$ ergs)  molecular 
outflow \citep{Garden:1991}.  In addition, 
an explosive, dispersal outflow is present in the DR21 Main clump \citep{Guzman:2024}. 
DR21 (OH) is another active, star forming clump within 
the DR21 cloud with multiple dense cores and molecular outflows \citep{Zapata:2012}. 
Such internal feedback activity may modify the local magnetic field geometry 
or increase the gas velocities to super-Alfv\'enic values that effectively reduces the impact of 
the magnetic field on the gas.

Previous studies of DR21 have explored the magnetic field strength and geometry.   \citet{Koley:2021} measured 
Zeeman splitting of the 1665 MHz and 1667 MHz OH lines toward the photodissociation region in DR21 Main.  From these data,
they calculate a line of sight magnetic field strength of 0.13 milli-gauss.  The mass to magnetic flux ratio is calculated assuming 
an abundance of OH molecules and assumed angles of the magnetic field direction with respect to the line of sight. 
The mass to magnetic flux ratios range from 6.6 to 1.1,  
which imply 
that the 
magnetic field can not support the DR21 Main clump against its own self-gravity. In addition, the magnetic energy density of the 
cloud 
is lower than the turbulent energy density by factors of 4 to 13 for this set of assumed angles.  From these results, it appears 
that the magnetic field is not dynamically important in this stage of the evolution of DR21 Main.

Polarization of dust grain emission at submm wavelengths over the DR21 cloud has been recently explored by \citet{Ching:2022} as part of the 
B-fields In STar-forming Region Observations (BISTRO) survey using SCUBA-2 on the James Clerk Maxwell Telescope.
They applied the Davis-Chandrasekhar-Fermi (DCF) method to estimate the magnetic field strength based on the angular dispersion of magnetic 
field angles in the plane of the sky, the volume density, and velocity dispersion \citep{Davis:1951,Chandrasekhar:1953}.  The resultant mass to 
flux ratios in the North and South areas of the DR21 cloud are 3.9 and 2.4 respectively.  They also found 
magnetic field orientations perpendicular to the orientation of the Main Ridge of the DR21 cloud for intermediate and high column densities. 

In this study, we investigate the role of the interstellar magnetic field in modulating the turbulent 
gas motions in the DR21 molecular cloud.  In Sect.~2, we briefly describe the 
data collection methods  and the resultant sensitivities.
In Sect.~3, we describe the Velocity Channel Gradient method that derives the position angle of 
molecular line brightness temperature gradients from which
the relative orientations with the 
the local magnetic field are evaluated
\citep{Lazarian:2018}.  The results of this analysis are presented 
in Sect.~4.  We discuss the implications of these results in Sect.~5.

\section{Data}
\subsection{Far infrared dust polarization}
Our science goals require high angular resolution measurements of the interstellar magnetic field topology and 
wide area coverage as provided by the {\it HAWC+} instrument 
on the Stratospheric Observatory for 
Infrared Astronomy 
(SOFIA).  
Far infrared polarization data were collected from {\it HAWC+}
for Project 09-0215 (P.I. T. Pillai) in September 2021 and 2022. 
Band E of the {\it HAWC+} instrument was used, which provided information at a central wavelength of 214$\mu$m.  The data 
were acquired using On-the-Fly (OTF) mapping mode using a scanning speed of 200 arcsec/sec. These data 
were processed using 
the {\it HAWC+} Data Reduction Pipeline version 3.2.1.dev0.
The angular resolution of the maps 
is 20.3\arcsec\ (half power beam width) and the pixel size is 4.55\arcsec.  The final product includes images of 
Stokes parameters I, U, and Q from which the degree of polarization, position angle of the interstellar magnetic field 
in the plane of the sky and random uncertainties are calculated. More complete descriptions of the data collection 
and reductions can be found in \citet{Pillai:2026}. 

\subsection{\co\ and \coa\ J=1-0 imaging}
The dust polarization data are complemented by \co\ and \coa\ J=1-0 spectral line data cubes from which gas motions are evaluated. 
\footnote{Data cubes of \co\ and \coa\ J=1-0 emission were downloaded from the Cygnus~X Survey data archive
at https://cygnus45.github.io/.}  Detailed descriptions of the dataset are 
summarized in \citet{Yamagishi:2018} and \citet{Takekoshi:2019}.
All spectral line data are smoothed to a half power beam width of 20.3\arcsec\ and an 
angular sampling interval of 4.55\arcsec\ to match the resolution and coverage of the {\it HAWC+} polarization data. 
The velocity resolution is 0.25~\kms.  The median root mean square (rms) values in main beam temperature units are 1.2~K (\co) and 
0.4~K (\coa) in 0.25~\kms\ channel widths.

\section{Description of the Velocity Channel Gradient Method}
Velocity channel gradient analysis (VChG) originally introduced by \citet{Lazarian:2018} is applied to 
\co\ and \coa\ J=1-0 emissions from DR21 using the implementation of this technique described 
by \citet{Heyer:2020}. 
VChG examines the 
relative orientations of the local magnetic field and the gradient of brightness temperatures in thinly sliced 
spectroscopic velocity channels.  
\citet{Lazarian:2000} demonstrated that the measured structure of antenna temperatures in 
velocity slices 
arises from velocity fluctuations when the 
observations satisfy the 
thin velocity slice condition ${\delta}v << \sigma_v(L)$ where ${\delta}v$ is the spectral 
channel width and $\sigma_v(L)$ 
is the velocity dispersion of the gas integrated over the size $L$ for the region of interest.
Integrating over the DR21 cloud ($L\sim$8~pc), the velocity dispersion is 2.1~\kms\ for \co\ and 
1.5~\kms\ for \coa.  For the Nobeyama CO data with 0.25~\kms\ wide channels, the thin-slice 
condition is 
satisfied. 

We define the $k$-th velocity slice of the data cube of
brightness temperatures $T(x,y,v_{\rm k})$ as $T_{\rm k}(x,y)$.  
Brightness temperature gradients are identified by partial derivatives 
along the x (R.A) and y (Declination) axes.
The gradient magnitude
and position angle for the $k$-th slice are
\begin{equation}
|\nabla{\rm T_k}(x,y)| =\bigg\{ \bigg (\frac{ \partial{ {\rm T_k(x,y)} } } {\partial{ {\rm x} } } \bigg )^2 +
                              \bigg (\frac{ \partial{ {\rm T_k(x,y)} } } {\partial{ {\rm y} } } \bigg )^2 \bigg\}^{1/2}
\end{equation}

\begin{equation}
\Psi_{\rm G,k}(x,y) = \arctan\left[\frac{-\partial{{\rm T_k(x,y)}}} {{\partial {\rm x}}} \middle/
                                   \frac{\partial{{\rm T_k(x,y)}}}  {{\partial {\rm y}}} \right]
\end{equation}
where 
$\Psi_{\rm G,k}$ is the position angle of the gradient measured counter-clockwise from the y (declination) axis that is identical 
to the definition of position angle applied to polarization measurements.
The gradients along the x and y axes 
 are calculated using the python function 
{\tt scipy.ndimage.filters.gaussian\_filter} using a kernel half-power beam width (HPBW) 
of 30.5\arcsec.
This kernel width is selected to minimize the effects of correlated data when calculating the gradients.
The returned gradients are rescaled by the pixel size in parsecs (0.031~pc/pixel) such that 
$\nabla_{\rm x} {\rm T}=\frac{ \partial{ {\rm T_k(x,y)} } } {\partial{ {\rm x} } }$ and 
$\nabla_{\rm y} {\rm T}=\frac{ \partial{ {\rm T_k(x,y)} } } {\partial{ {\rm y} } }$ are 
in units of K/pc.  
In order to be consistent with \citet{Lazarian:2018}, 
we do not rotate $\Psi_{\rm G,k}$ by 90$^\circ$ as was done by \citet{Heyer:2020} and 
\citet{Pillai:2026} 
to compare with the local 
magnetic field orientation. 

In this study, our calculation of gradients differs from that of \citet{Lazarian:2018} and \citet{Hu:2020}.
They calculate gradients and orientations for all pixels within a sub-block to construct a distribution of 
gradient angles.  The most probable angle and its uncertainty for the sub-block 
are determined from a Gaussian fit to the distribution. 
The size of a sub-block is set by the resultant angle uncertainty that ensures a reliable gradient angle. 
In contrast, we calculate gradients within each velocity slice over distances of 30.5\arcsec\ 
 that is 
set by the convolving kernel.  This allows us to exploit the high angular resolution of the spectral 
line and magnetic field data while minimizing the effect of correlated noise. 

Uncertainties in the derived gradient values due to the noise in the observations 
are derived for each axis using the following method. 
First, for each axis, gradients, $\nabla_{\rm x} {\rm T_N}$, $\nabla_{\rm y} {\rm T_N}$, are calculated for all positions 
in velocity channels 
with no expected line emission from the DR21 and W75 clouds or the CO outflow. 
Gaussian profiles are separately fit to the composite distributions of $\nabla_{\rm x} {\rm T_N}$ 
and $\nabla_{\rm y} {\rm T_N}$. 
The standard deviations of the gaussian fits represent the average gradient uncertainties for each axis.
The measured uncertainties of the antenna temperature gradients for the x and y axes are
5.26 and 5.31~K pc$^{-1}$ for \co\ and 2.23 and 2.25 ~K pc$^{-1}$ for \coa. 
The uncertainties of the total gradient $|\nabla{\rm T_k(x,y)}|$ and angular direction of the gradient, 
$\Psi_{\rm G,k}$ are
\begin{equation}
\sigma(|\nabla{\rm T_k(x,y)|})=\frac{
\left [(\nabla_{\rm x} {\rm T})^2(\sigma(\nabla_{\rm x} {\rm T_N})^2+
       (\nabla_{\rm y} {\rm T})^2(\sigma(\nabla_{\rm y} {\rm T_N})^2\right]^{1/2}}
       {\nabla{\rm T_k(x,y)}}
\end{equation}

\begin{equation}
\sigma({\rm \Psi_{\rm G,k}(x,y)})=\frac{
       (\nabla_{\rm x}{\rm T}/\nabla_{\rm y}{\rm T})}
      {(1+(\nabla_{\rm x}{\rm T}/\nabla_{\rm y}{\rm T})^2)}
\left [\left (\frac{\sigma(\nabla_{\rm x}{\rm T_N})}{\nabla_{\rm x}{\rm T}}\right )^2+
       \left (\frac{\sigma(\nabla_{\rm y}{\rm T_N})}{\nabla_{\rm y}{\rm T}}\right )^2\right]^{1/2}
\end{equation}

The primary product of the gradient analysis is
a set of angles, $\Psi_{\rm G,k}$,  that can be compared to the magnetic field orientation derived 
from dust polarization at the same position.
The difference between angles is 
\begin{equation}
\Phi_{\rm k}(x,y)=\Psi_{\rm G,k}(x,y)-\Psi_{\rm B}(x,y)
\end{equation}
where
$\Psi_{\rm B}(x,y)$ is the position angle of the interstellar magnetic field component
in the plane of the sky at position (x,y) derived from far infrared polarization measurements.
The uncertainty of the relative angle between the magnetic field and the gradient is 
\begin{equation}
\sigma(\Phi_{\rm k}(x,y))=\left [ \sigma^2(\Psi_{\rm G,k}(x,y)) +
                           \sigma^2(\Psi_{\rm B}(x,y)) \right]^{1/2}
\end{equation}

The projected Rayleigh statistic, \PRS,  is a quantitative measure that describes the degree of alignment
between 2 sets of angles.
It is defined as
\begin{equation}
{\rm V}_{\rm k} =\frac{\sum_{i=1}^{{\rm n}_{\rm ind}} \cos(2\Phi_{\rm k}(x_i,y_i))}{({\rm n}_{\rm ind}/2)^{1/2}} 
\end{equation}
where ${\rm n}_{\rm ind}$ is the number of independent pixels in the selected area \citep{Jow:2018}.   The 
pixel gridding for the {\it HAWC+} image of magnetic field position
angles is oversampled by a factor of $\sim$4.5. Since the Nobeyama data are matched to the resolution and pixel gridding 
of the {\it HAWC+} image, the gradients and gradient angles  are similarly oversampled.  
For oversampled data, calculating ${\rm V}_{\rm k}$ by summing over all pixels, ${\rm n}_{\rm {pix}}$  rather 
than ${\rm n}_{\rm ind}$ 
can introduce an additive bias \citep{Fissel:2019}. 
To correct for this bias, 
we follow the method described by \citet{Fissel:2019}.  First, we calculate the ${\rm V}_{\rm k}^\prime$ using all 
pixels within the target area, 
\begin{equation}
{\rm V}_{\rm k}^\prime =\frac{\sum_{i=1}^{{\rm n}_{\rm pix}} \cos(2\Phi_{\rm k}(x_i,y_i)) } { ({\rm n}_{\rm pix}/2)^{1/2}}.
\end{equation}
For a given velocity slice, ${\rm k}$, we construct a noise slice of antenna temperatures, $T_{\rm k,Noise}$ 
by random sampling a normal distribution whose 
standard deviation is equal to the derived rms value at each pixel.  These data are further convolved to 
the same resolution as the CO data.  Gradients of antenna temperature 
and relative orientations of the gradient angle with respect to the magnetic field 
are calculated using the same kernel that is applied to the CO data. 
The resultant $\phi_{\rm k,Noise}$ values are used to calculate 
${\rm V}_{\rm k,Noise}^\prime$ as in equation~8.
These steps are repeated 256 times 
to build a distribution of ${\rm V}_{\rm k,Noise}^\prime$ values.  The variance of this distribution, 
$\sigma({\rm V}_{\rm k,Noise}^\prime)^2$ 
is an estimate of the oversampling factor, ${\rm n_{pix}}/{\rm n_{ind}}$.  The \PRS\ value corrected for oversampling is
\begin{equation}
{\rm V}_{\rm k}={\rm V}_{\rm k}^\prime/\sigma({\rm V}_{\rm k,Noise}^\prime)
\end{equation}
This is repeated for all spectral channels in the \co\ and \coa\ J=1-0 data between $-$30 and 30~\kms. 
Following this correction for oversampling, the standard deviation of ${\rm V}$ values over spectral channels with no 
CO emission is $\sim$1 \citep{Fissel:2019}.
Angle sets with ${\rm V}_{\rm k} >> 0$ are interpreted as strongly aligned, parallel features.
Sets with ${\rm V}_{\rm k} << 0$ indicate features that are strongly orthogonally oriented.
Values of ${\rm V}_{\rm k} \sim 0$ can result from random,
uniformly distributed orientations or orientations clustered near 45 degrees.
The null hypothesis is that the angle distribution is uniform and can be 
rejected at probabilities of 5\% and 0.5\% for $|$V$|$ values
$>$ 1.64 and $>$2.57 respectively \citep{Batschelet:1972}. 
A 3$\sigma$ confidence level corresponds approximately to $|{\rm V}|\sim$2.87 \citep{Soler:2021}. 

\section{Results}
\begin{figure*}[htb]
\begin{center}
\includegraphics[width=0.9\textwidth,trim={0 0 0 0}, clip]{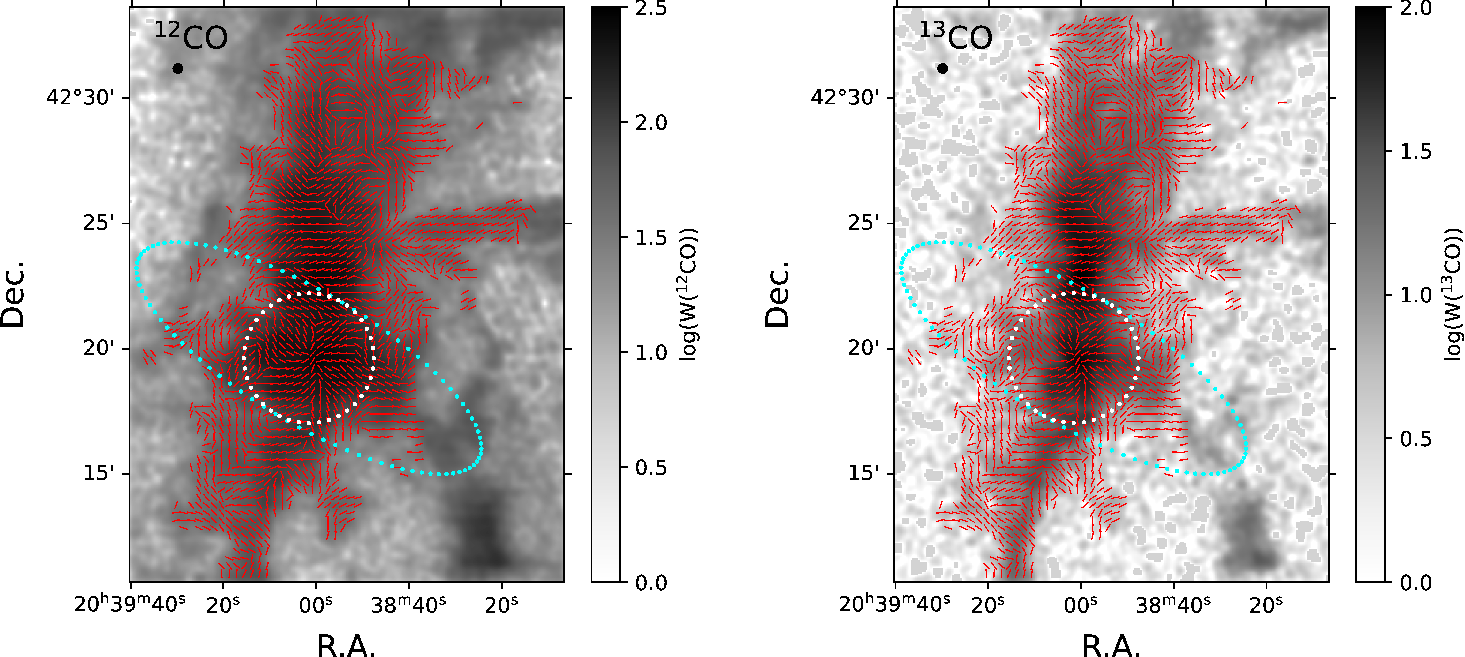}
\caption{Images of velocity integrated \co\ (left) and \coa\ (right) J=1-0 emission from the DR21 cloud.
The intensity stretch is logarithmic to highlight the faint, surrounding features with respect to 
the bright emission from the Main Ridge. The geometry of the magnetic field component in the plane of the sky is 
represented by equal-length red segments sampled every 4$^{th}$ pixel (every 18.2\arcsec) for clarity.  The cyan ellipse 
shows the extent and orientation of the CO molecular outflow \citep{Garden:1991}. The white circle 
highlights the area where there is a quadrupole-like  
distribution of magnetic field position angles. The filled, black circle represents the beam size.
}
\label{figure1}
\end{center}
\end{figure*}

The synthesis of spectroscopic data of molecular line emission and dust polarization measurements
offers important 
insight to the gas kinematics of molecular clouds. 
Figure~\ref{figure1} shows magnetic field orientations in the plane of the sky and velocity integrated 
intensity images of \co\ and \coa\ J=1-0 emission 
over the velocity interval [$-$7,1.5]~\kms\ that covers most of the emission from DR21 but excludes 
the high velocity wings of the 
molecular outflow \citep{Garden:1991}.  
The magnetic field orientations in the plane of the sky are represented by equal length, red segments plotted 
every 4th pixel on the x and y axes. 
Most of the molecular line emission is from the bright, 
Main Ridge of the cloud elongated along the North-South direction.  It includes the two primary star formation sites DR21 Main and DR21 OH.  
This Main Ridge is surrounded by 
several faint, narrow sub-filaments 
that have been identified in previous studies of dust and line emission \citep{Hennemann:2012}. 

The magnetic field orientations exhibit changing patterns as one surveys from west to east across the cloud. 
Along the northern 
edges of the Main Ridge between declinations 42$^\circ$25$^m$ and 42$^\circ$30$^m$, 
the magnetic field segments have position angles of $\sim$90$^\circ$ that flip to $\sim$0$^\circ$ in the 
interior regions and then $\sim$45$^\circ$ along the eastern cloud edge.  
Further south, the magnetic field segments exhibit a quadrupole-like orientation 
centered 
on the DR21 Main star forming core (white circle in Figure~1). Two opposing poles have magnetic field position angles ranging from 30$^\circ$ to 55$^{\circ}$
that 
are aligned with the molecular outflow indicated by the cyan colored ellipse in Figure~1.
The other two poles show magnetic field position angles approximately orthogonal to the outflow 
that may indicate locally collapsing motions that are dragging the magnetic field along 
these polar directions \citep{Lazarian:2018}.  Other locations show swirling patterns of magnetic field segments within the 
covered area.

\begin{figure*}[htb]
\centering
\begin{tabular}{l c r}
\includegraphics[width=0.45\textwidth,trim={0 0 0 0},clip]{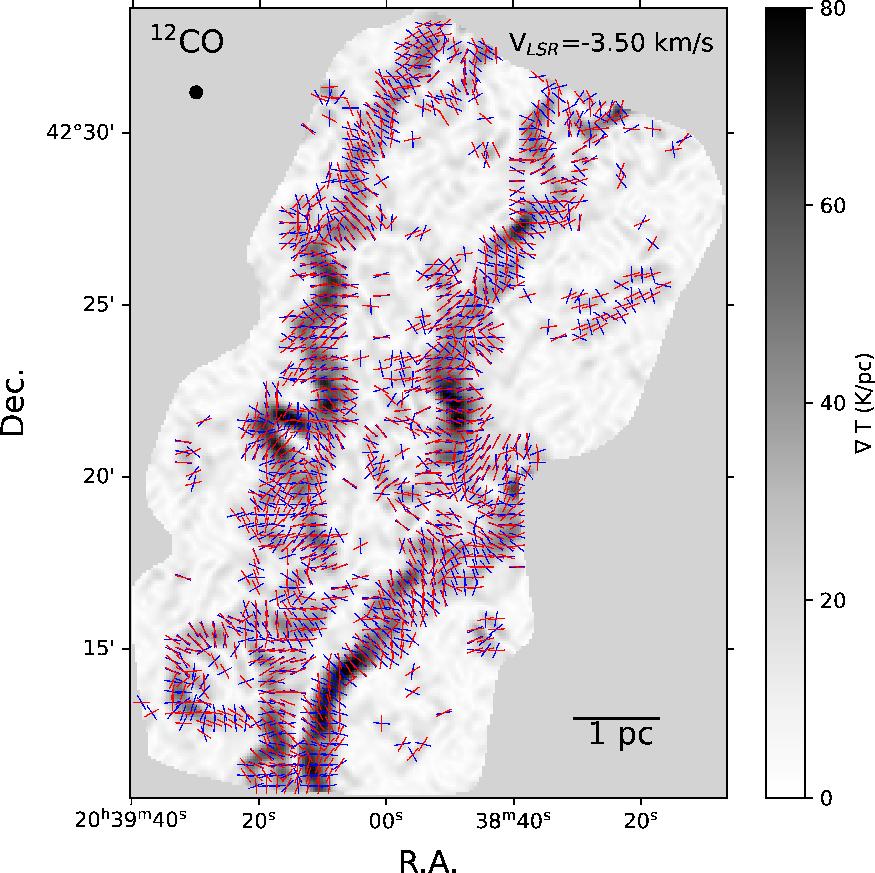} 
&
&
\includegraphics[width=0.45\textwidth,trim={0 0 0 0},clip]{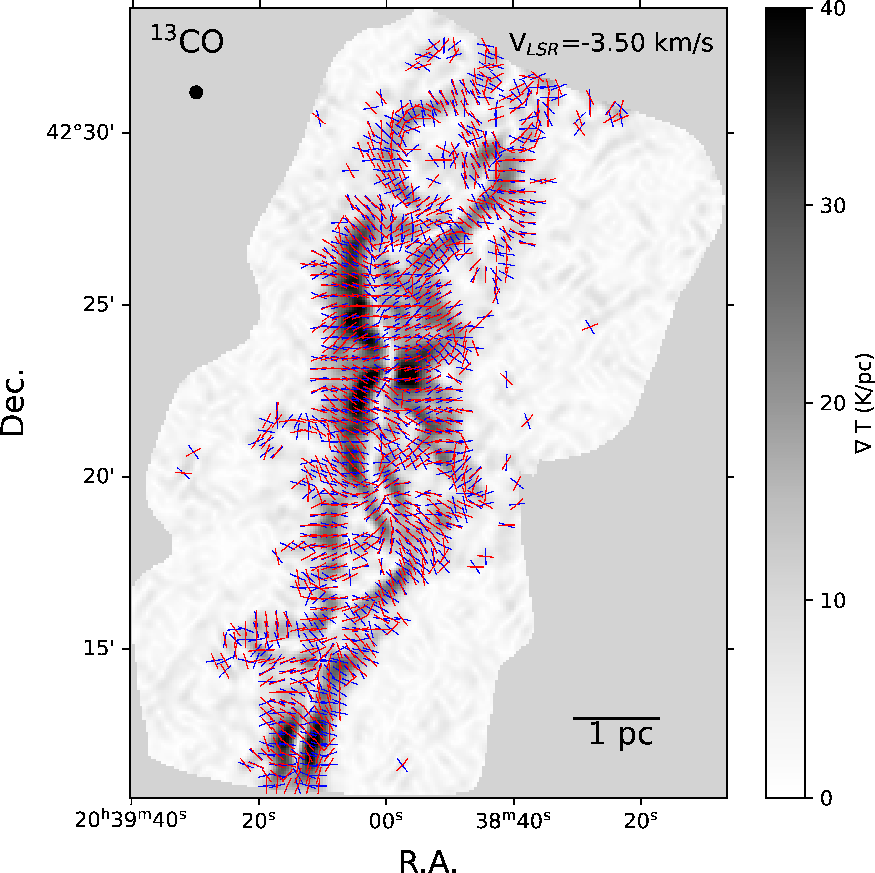} 
\end{tabular}
\caption{Images of $\nabla T$, 
derived from (left) \co\ emission and (right) \coa\ emission for \vlsr=$-$3.5~\kms. 
The blue segments mark the derived gradient position angle, $\Psi_{\rm G,k}$
for pixels where $|\nabla T|/\sigma(|\nabla T|) \ge 3.5$. 
Red segments show the position angle of the plane of the sky 
component of the magnetic field.  
For clarity, the segments are drawn every 4 pixels.
}
\label{figure2}
\end{figure*}

The plane of the sky gradient of antenna temperatures are calculated for  
each 0.25~\kms\ 
wide velocity channel over the velocity range [$-$30,30]~\kms\ for \co\ and 
\coa.  
Figure~\ref{figure2} shows images of the derived 
gradients of \co\  and \coa\ 
brightness temperatures within the velocity channel of width 0.25~\kms\ at velocity $-$3.5~\kms. 
Blue line segments represent the orientations of the gradient vector.
The drawn line segments are restricted to those positions with $|\nabla T|/\sigma(\nabla T) > 3.5$ to ensure robust 
values of $\Psi_{\rm G,k}$ are displayed.
This signal to noise condition corresponds to an uncertainty in $\Psi_{\rm G,k}$ values less than 15$^\circ$.
In the northern part of the Main Ridge ($\delta > 42^\circ 25^m$) and for most velocity channels, 
the \co\ emission is smoothly distributed so 
the interior positions do not satisfy this signal to noise ratio, which indicates no significant gradients in brightness temperatures.  
This flat distribution of \co\ antenna temperatures is likely a result of the high optical 
depth of the \co\ emission in which the line is saturated. 

Flat distributions in \coa\ velocity channel images are limited to smaller areas along the spine of the DR21 cloud, where one might 
expect higher line opacities. To evaluate 
whether opacity can significantly impact our analysis, we derive the relative orientations between \coa\ integrated intensity and molecular 
hydrogen column density from dust continuum observations with the {\it Herschel Observatory} \citep{Pokhrel:2020}.  The velocity interval is 
[-7.0,1.5]~\kms\ that covers most of the emission from the DR21 cloud.  The image of \coa\ integrated intensity 
gradients and orientation segments for both \coa\ and \htwo\ column density are shown in Figure~{A.1} in the Appendix.  For clarity, 
segments are only drawn for pixels with gradient magnitude signal to noise ratio greater than 3.5.  
Using all pixels and applying a correction for oversampling, 
the projected Rayleigh statistic is 63, which indicates strong parallel alignment between the gradient orientations for these 
mostly optically thin gas tracers. The overall low optical depth of \coa\ emission limits the impact of line opacity on the gradient 
orientation. 

Photodissociation can also create brightness temperature gradients that are independent of MHD turbulence.  
At the cloud edges, the CO abundance is reduced as the far-ultraviolet radiation field from a nearby, massive star or the ambient 
FUV radiation field dissociates CO molecules
\citep{Hollenbach:1997}.
The reduced 
CO column density leads to 
lower brightness temperatures relative to those produced by more abundant CO molecules in nearby, more self-shielded regions, 
which creates a gradient.
However, gradient orientations induced by photodissociation would necessarily be pointed toward local, bright  FUV sources and 
not along the magnetic field direction.  FUV sources are clustered within the DR21 Main star forming region 
but there are no strong, localized alignments of \co\ or \coa\ brightness temperature gradient orientations toward these massive stars. 
We conclude that the photodissociation process is not responsible for the observed brightness temperature gradients in DR21.

\subsection{Rayleigh statistic over the full cloud}
\begin{figure*}[htb]
 \centering
\includegraphics[width=0.8\textwidth,trim={0 0 0 0},clip]{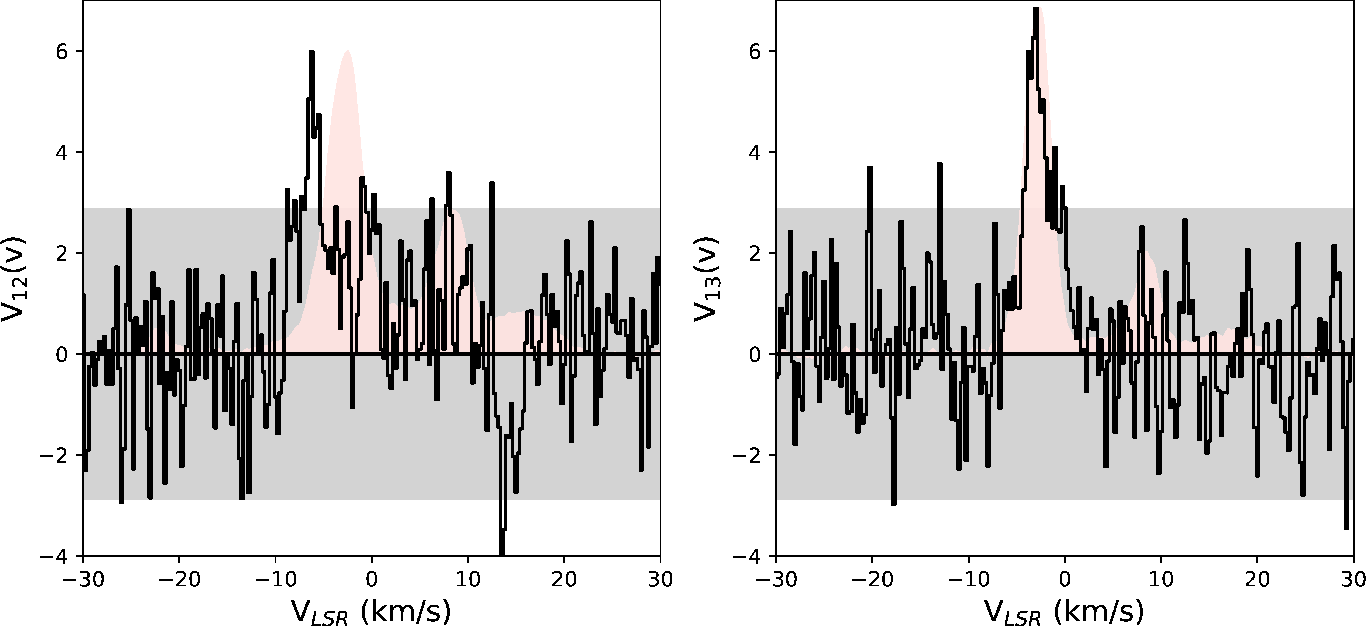} 
\caption{Projected Rayleigh statistic spectra for (left) \co\ and (right) \coa\ derived over the full area of the DR21 cloud 
(black).  The red shading in each panel shows the mean \co\ and \coa\ spectra over the whole map.  These spectra
have been scaled so that the peak temperature matches the peak \PRS\ value within the DR21 velocity interval.
The grey shading marks $|{\rm V}| \le 2.87$ that approximates the 3$\sigma$ or less 
confidence level for the projected Rayleigh statistic. 
}
\label{figure3}
\end{figure*}

The projected Rayleigh statistic is a simple, quantitative metric to determine 
whether the set of relative orientations is mostly perpendicular or mostly parallel or mostly
 random for a given velocity slice.  
The spectra of the projected Rayleigh statistic for \co\ and \coa\ are 
shown in Figure~\ref{figure3}.  
The grey shaded region shows the $\pm$2.87 threshold
that corresponds to $<$3$\sigma$ confidence level \citep{Soler:2021}.  
The red shaded area in each panel shows the mean \co\ and \coa\ spectrum averaged 
over the full area of the cloud to provide context between the line emission and the 
gradients.  
The \co\ \PRS\ shows a sequence of 
positive values between $-$8~and+1~\kms (DR21).

The dip in the \co\ \PRS\ values between -5 and -2 \kms\ is a result of the wide 
area of saturated line emission that does not generate significant brightness temperature gradients.  
The gradient angles from these areas are randomly distributed, which leads to a decrease in the PRS values
towards zero 
over this velocity  interval. 
\coa\, positive \PRS\ values range between $-$6~and~+1~\kms\ closely matching the 
range of \coa\ emission defined by the red shaded area.  There is no dip in the \coa\ PRS values
as the \coa\ emission is mostly optically thin and not saturated like the \co\ emission. 
Overall, the positive \PRS\ values derived from \co\ and \coa\ data point 
towards a preferred parallel alignment between the magnetic field orientation and $\Psi_{G}$ in the DR21 
molecular cloud.

\subsection{Rayleigh Statistic in regions of interest}
The DR21 molecular cloud presents a very complex environment given the dominant 
Main Ridge, smaller sub-filaments, several sources of energy 
feedback from recently formed stars, and possible interactions with the W75 cloud. 
Given this complexity, calculating the projected Rayleigh 
statistic for each velocity slice over the full cloud may dilute the value as one environment may cancel the 
contributions of another region or simply be biased by the Main Ridge, which comprises 
most of the mass and area of the DR21 cloud. 

\begin{figure}[htb]
\centering
\includegraphics[width=0.4\textwidth,trim={0 0 0 0},clip]{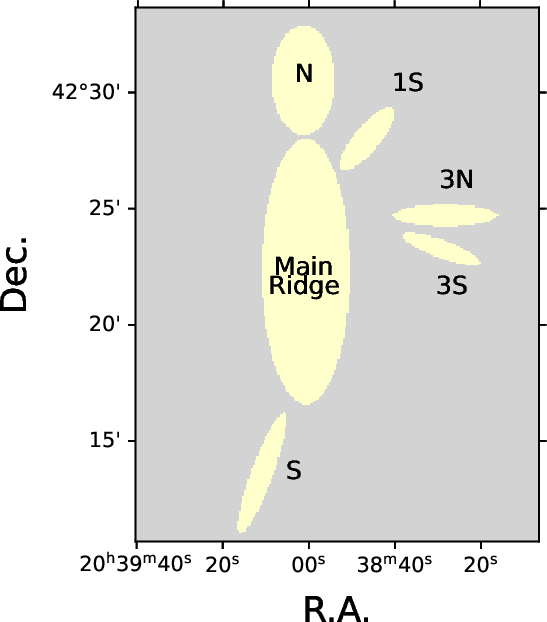} 
\caption{Selected regions of interest in the DR21 cloud.}
\label{figure4}
\end{figure}

In this section, we explore the relative orientations between antenna temperature gradients and the plane 
of the sky magnetic field component in several local regions of interest within the DR21 cloud 
previously defined by 
\citet{Pillai:2026} 
(see their Figure~3) 
but excludes the region of interest that defines 
the molecular outflow.  
Figure~\ref{figure4} shows the image mask
for the 6 regions of interest that are analyzed.   
The list includes the Main Ridge  
and five appending, narrow sub-filaments that surround the cloud (N, 1S, 3N, 3S, S).  
The 5 sub-filaments have been previously 
identified in {\it Herschel} data  by \citet{Hennemann:2012} as N, F1S, F3N, F3S, and S. 
Each region of interest is represented by an ellipse 
that describes its major and minor axes and center position.  
The ellipse provides a mask that allows one to isolate all pixels within 
the area of the ellipse to derive local values of the projected 
Rayleigh statistic.  The velocity dispersions for the 5 sub-filaments 
range from 0.53~\kms\ (3S) to 1.6~\kms\ (N).  The thin-slice condition is only marginally satisfied 
for 
sub-filament 3S.

\begin{figure*}[htb]
\centering
\begin{tabular}{l c r}
\includegraphics[width=0.45\textwidth,trim={0 0 0 0},clip]{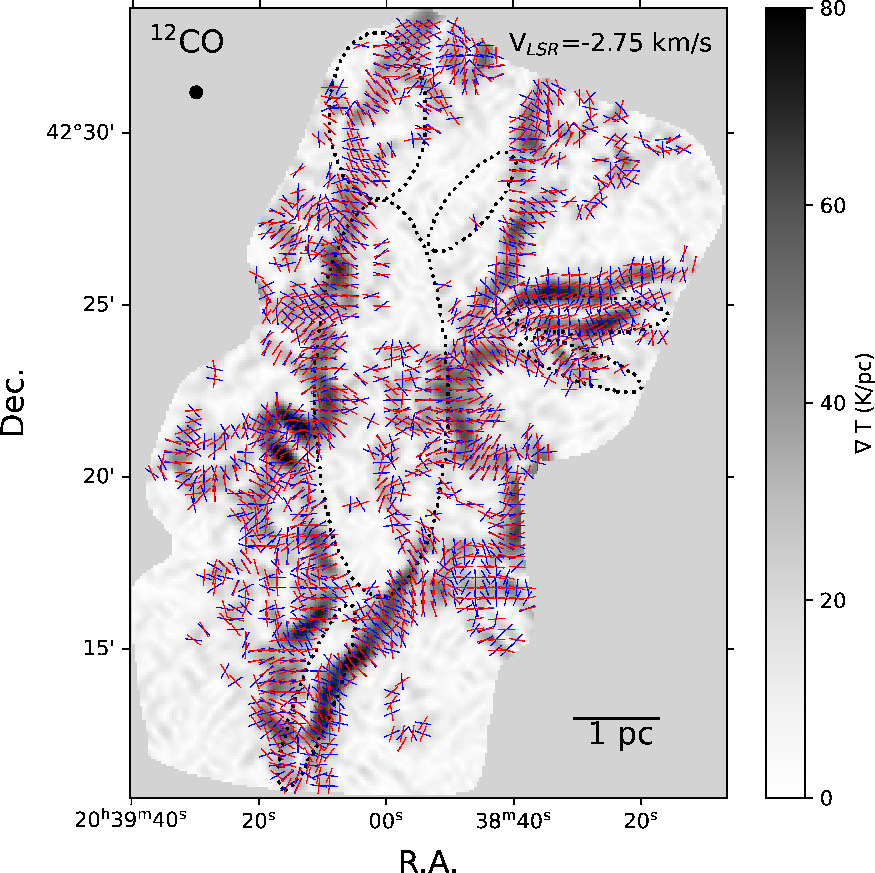}
&
&
\includegraphics[width=0.45\textwidth,trim={0 0 0 0},clip]{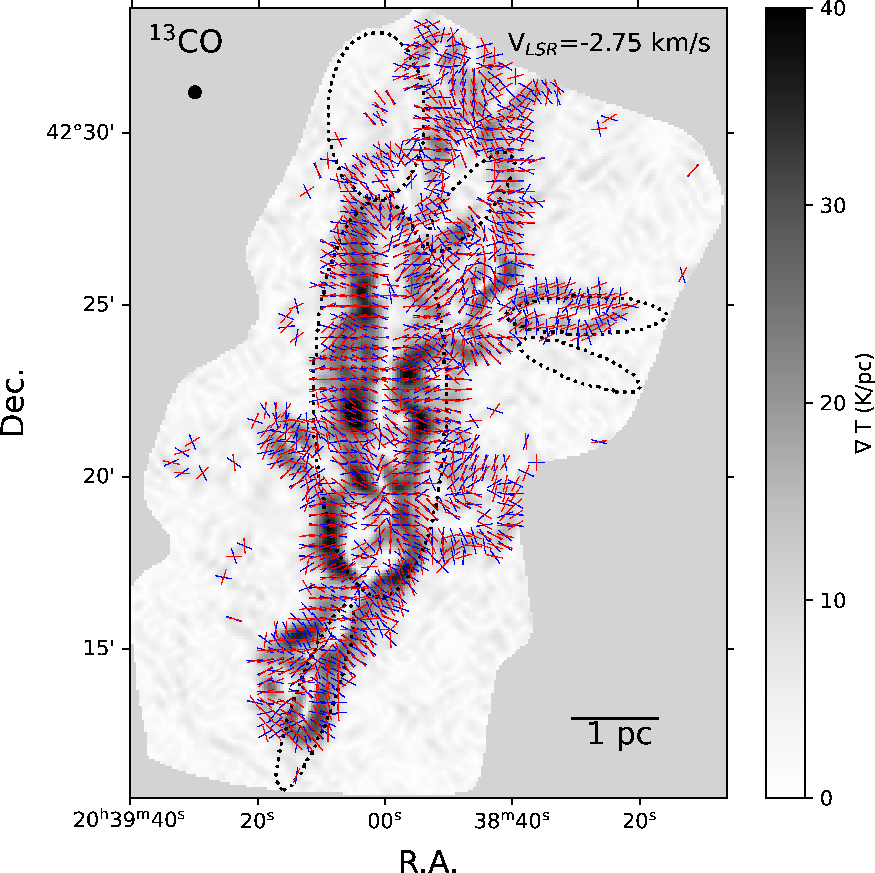}
\end{tabular}
\caption{(left) Images of $\nabla T$,
derived from (left) \co\ emission and (right) \coa\ emission for \vlsr\ $-$2.75~\kms.
The blue segments mark the derived gradient position angle, $\Psi_{\rm G,k}$
for pixels where $|\nabla T|/\sigma(|\nabla T) \ge 3.5$.  The black, dotted line ellipses 
outline the area of each region of interest.
Red segments show the position angle of the plane of the sky
component of the magnetic field.  
For clarity, the segments are drawn every 4 pixels.
}
\label{figure5}
\end{figure*}

\begin{figure*}[htb]
\centering
\includegraphics[width=0.9\textwidth,trim={0 0 0 0},clip]{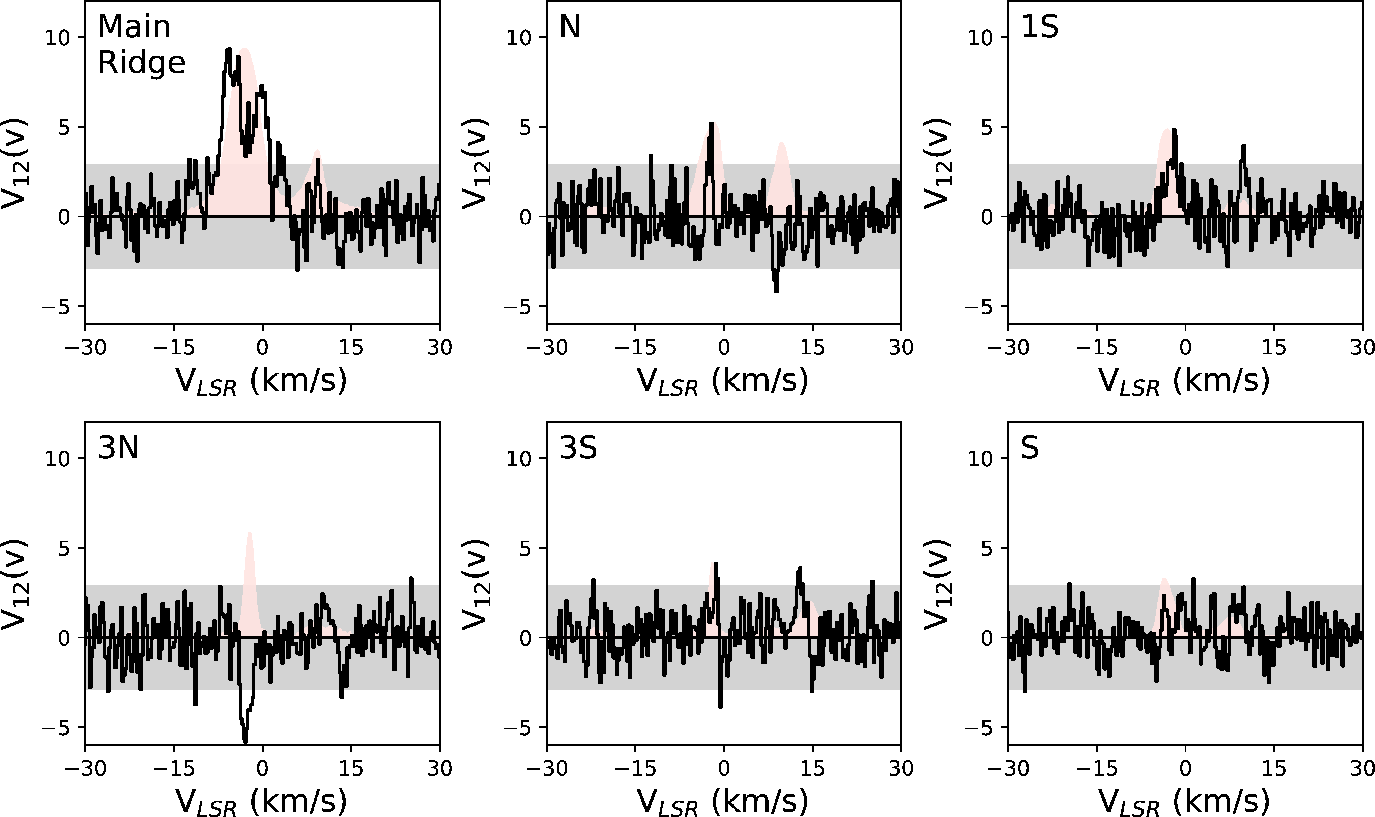} 
\caption{Projected Rayleigh statistic spectra (black) for \co\ for each region of interest.
The red shading is the \co\ mean spectra over each region of interest area.
These spectra
have been scaled so that the peak temperature matches the peak absolute value of the \PRS\ 
within the DR21 velocity interval.
The filament label 
is at the top left of each plot. }
\label{figure6}
\end{figure*}

\begin{figure*}[htb]
\centering
\includegraphics[width=0.9\textwidth,trim={0 0 0 0},clip]{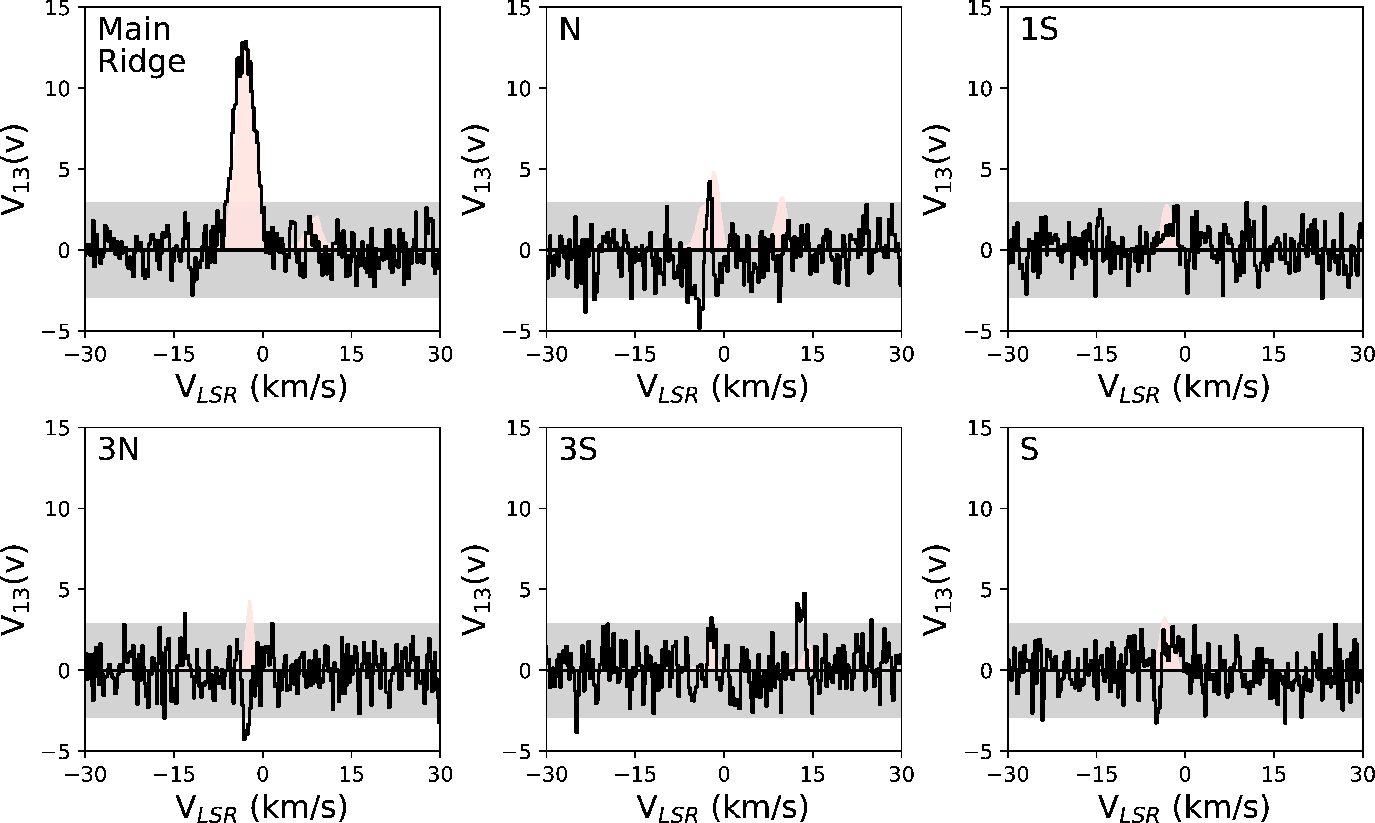} 
\caption{Projected Rayleigh statistic spectra for \coa\ for each region of interest. 
The red shading is the \coa\ mean spectra over each region of interest area.
These spectra
have been scaled so that the peak temperature matches the peak absolute value of the \PRS\ within the DR21 velocity interval.
}
\label{figure7}
\end{figure*}

Figure~\ref{figure5} shows images of the derived gradients of
\co\ and \coa\ brightness temperatures within the velocity
channel at $-$2.75~\kms, in which several of the sub-filaments (N, 3N, 3S, S) are emitting. 
The black dotted ellipses show the area of each sub-filament. 
The projected 
Rayleigh statistic spectra for each region of interest are displayed in Figure~\ref{figure6} for \co\ 
and Figure~\ref{figure7} for \coa.
In the Main Ridge, \ZCO\ shows a strong, positive, triple peaked profile 
over the velocity interval of the cloud.
The \ZCOA\ Main Ridge spectrum also presents a strong, positive feature over this velocity interval. 
In the N filament,  \ZCO\ and \ZCOA\ exhibit both positive and negative features within 
the velocity interval [$-$6,1]~\kms\ while \ZCO\ also shows a weak, negative feature within the 
velocity interval of W75 (8-11~\kms).  Significant positive values of the Rayleigh statistic in both \co\ and \coa\ 
are 
also found in the 1S filament. The 3S filament includes a weak, positive component at \vlsr=$-$1~\kms and a component 
at 
\vlsr=13.5~\kms\ in both \co\ and \coa, which is likely from 
another cloud along the line sight that is not related to the Cygnus star forming complex. 
The 3N filament shows the only unambiguous negative features in \ZCO\ and \ZCOA\ within the DR21 velocity interval.  
In the S filament, sequences of weak, positive \PRS\ features are identified in the velocity intervals of both the DR21 and W75. 

Multiple velocity components in \co\ and \coa\ spectra over the observed 
field provide evidence for the superposition of molecular clouds along the line of 
sight. These components appear in velocity intervals
[-7,1.5]~\kms\ (DR21), [5,11]~\kms\ (W75), and [12,18.5]~\kms\ (unnamed cloud).
These secondary components also appear within \PRS\ spectra derived from sub-filaments
N, 1S, 3N, and 3S.  

While the velocity components can be isolated when deriving 
brightness temperature gradient magnitudes and orientations, the U and Q polarization 
measurements 
are weighted by dust column density 
along the line of sight.  This weighting impacts the polarization fraction and orientation 
 and can not be readily isolated along the spectral axis. 

To evaluate if the magnetic field orientations in the sub-filaments 
are affected 
by the superposition of the 3 clouds, we have calculated the 
\htwo\ column densities from the \coa\ integrated 
intensities for each velocity interval and each sub-filament as a proxy for dust 
column density.  \coa\ column density, ${\rm N}_{13{\rm CO}}$ is derived from the expression,
\begin{equation}
{\rm N}_{13{\rm CO}}=\frac{8\pi k_B \nu^2}{hc^3A_{10}} \biggl(\frac{Z}{2J+1}\biggr) 
exp\biggl(\frac{hB_\circ J(J+1)}{k_B T_{ex}}\biggr)\int T_{mb}(v)dv \; cm^{-2}
\end{equation}
where $k_B$ is the Boltzmann constant, c is the speed of light, $A_{10}$ is the Einstein spontaneous 
rate coefficient for the \coa\ J=1-0 transition, $B_\circ$ is the rotational constant,
and Z is the partition function.
This expression for ${\rm N}_{13{\rm CO}}$ assumes optically thin \coa\ emission, 
local thermodynamic equilibrium (LTE), and 
$T_{ex} >> hB_\circ/k_B$. 
The \htwo\ column densities are 
derived by 
\begin{equation}
{\rm N}({\rm H_2})=[{\rm H_2}/^{13}{\rm CO}] {\rm N}_{13{\rm CO}}. 
\end{equation}
Assuming an excitation temperature of 30~K, a 
relative \htwo\ to \coa\ abundance ratio of 5$\times$10$^{5}$ and 
Z$\approx$$k_B T_{ex}/hB_\circ$,  we calculate \htwo\ column densities for all pixels 
within each sub-filament mask and for each velocity interval. 
The median \htwo\ column \htwo\ densities
are listed in Table~\ref{tab:table1}. 

The DR21 cloud contributes the largest fraction of the total column density in all 
sub-filaments. 
The DR21 median column densities are 2 to 4 times larger than the W75 values and much larger than 
the unnamed cloud values.  
From these column density comparisons, the 
observed magnetic field orientations mostly reflect those associated with the DR21 sub-filaments and 
that the superposition of clouds does not strongly impact the derived relative orientations of the DR21 cloud.

\begin{table*}
\centering
\caption{Median \htwo\ column densities for each velocity interval within each sub-filament}
\begin{tabular}{lccc}
\hline
\hline
Sub-filament &  & log(N(\htwo) cm$^{-2}$) &  \\
Cloud & DR21 & W75 & Unnamed \\
Velocity Interval (\kms)  & [-7,1.5] & [5,11] & [12,18.5] \\
\hline
 Main Ridge  &  21.65 & 21.25 & 20.78\\
 N  &  21.68 & 21.19 & 20.68 \\
1S  &  21.55 & 21.15 & 20.69 \\
3N  &  21.71 & 21.10 & 20.72 \\
3S  &  21.56 & 21.22 & 20.89 \\
S   &  21.35 & 21.14 & 20.85 \\
\hline
\smallskip
\end{tabular}
\label{tab:table1}
\end{table*}

\section{Discussion}
The theory of MHD anisotropy in the interstellar medium has been developed by \citet{Goldreich:1995} and \citet{Lazarian:1999}.   Both studies conclude 
that velocity anisotropy results from the interaction between the 
magnetic field 
and  turbulent gas motions.  This kinematic anisotropy is 
realized in the form of elongated turbulent gas eddies that naturally develop in a magnetized medium. 
Computational simulations with 
varying values of $\beta=2({\rm M_A}/{\rm M_s})^2$, where ${\rm M_A}$ and ${\rm M_s}$ are 
the Alf\'enic and thermal Mach numbers respectively, 
demonstrate 
induced anisotropy for each propagating linear eigenmode (Alfv\'en, slow and fast modes) \citep{Lazarian:2018}. 
Each mode is a solution to the MHD equations within a medium with constant density and 
pressure that is threaded 
by a magnetic field.  
Alfv\'en waves generate incompressible perturbations that are perpendicular to the wave vector
direction. 
The properties of the compressive slow and fast modes depend upon the relative orientation between the wave vector and 
magnetic field and whether the plasma is in a weak field (sound speed greater than \VA) or 
strong field environment (sound speed less than \VA).
For Alfv\'en and slow modes,  \citet{Lazarian:2018} find the long axes of the MHD eddies are parallel to 
the local magnetic field orientation while the long axes of fast mode eddies are perpendicular 
to this orientation. 
These modes can also interact with each other.  For example, Alfv\'en waves can 
shear slow modes such that the slow modes inherit some of the properties of the 
Alfv\'en modes \citep{Goldreich:1995, Cho:2003}. 

\citet{Lazarian:2018} 
introduced the Velocity Channel 
Gradient method applied to 
spectroscopic data cubes of brightness temperature 
as an observational tool to investigate this induced anisotropy.
They constructed thin-slice velocity channel intensity images (see their equation~5) from the MHD simulations
for each MHD mode. 
From these images, 
gradients of intensity and gradient position angles were calculated and treated as measures of 
velocity fluctuations linked 
to MHD eddies.
For all MHD modes (Alfv\'en, slow, fast), they compared the gradient orientation 
to the magnetic field orientation.  The theoretical predictions for each mode 
were verified in their forward modeled simulations. 
Gradient angles were found to be perpendicular to the magnetic field orientations 
for Alfv\'en and slow modes and parallel to the magnetic field orientations 
for fast modes.
In the case of a self-gravitational collapsing cloud, the gradient angles are expected to 
be parallel to the magnetic field \citep{Lazarian:2018, Hu:2020}. 

\subsection{DR21 Main Ridge}
Velocity Channel Gradient analysis of \co\ and \coa\ J=1-0 emission from the DR21 cloud 
identifies both parallel and perpendicular alignments of the antenna temperature 
gradient angle, $\Psi_{\rm G,k}$,  with the magnetic field orientation 
in the plane of the sky.   These alignments are apparent 
quantitatively in the projected Rayleigh statistics
over the velocity intervals of the cloud (see Figure~\ref{figure3}, Figure~\ref{figure6}, and 
Figure~\ref{figure7}).   
Overall, there is much more area in the cloud where $\Psi_{\rm G,k}$ values are approximately parallel to the magnetic field. This is most visible 
along the Main Ridge in the \coa\ data.  

The parallel orientations between $\Psi_{\rm G,k}$ and $\Psi_{\rm B}$ can result from two 
different physical processes -- strong self-gravity in a locally collapsing area of the cloud 
and dominant MHD fast modes.
In the absence of strong self-gravity and a low $\beta$ environment, the simulation results of 
\citet{Lazarian:2018} demonstrate that MHD fast modes can also lead to parallel alignment between gradient and magnetic field 
position angles as measured in our analysis.  This alignment results from the long axis of the elongated eddies being 
perpendicular to the magnetic field, which leads to intensity 
gradients parallel to the magnetic field orientations. 
In high $\beta$ environments, fast modes lead to isotropic distributions of 
velocity \citep{Cho:2003, Lazarian:2018}. 

The role of gravity in modulating the relative orientations between gradients and the magnetic field 
has been 
well established from both observations \citep{Planck:2016,Soler:2017a} and theory \citep{Soler:2013,Soler:2017b,Lazarian:2018, Hu:2020, Seifried:2020}.
Specifically, the relative orientations between the gradient and magnetic field are found to transition  
from mostly perpendicular in low column density regions of a cloud to 
mostly parallel in high column density regions \citep{Planck:2016, Soler:2017a}. 
When the gravitational energy density of an 
eddy whose long axis is aligned with the magnetic field 
exceeds the
MHD turbulent pressure, gas is accelerated along the magnetic field lines leading to velocity gradients that are 
parallel to the magnetic field \citep{Hu:2020}.  
Parallel alignments can also result from compressive motions within the cloud 
due to the gravitational collapse of the cloud itself or converging flows \citep{Soler:2017b}.
This parallel condition is present in most of the Main Ridge of the DR21 cloud.

Direct evidence for infall motions in the DR21 cloud is provided by \citet{Schneider:2010} and \citet{Bonne:2023}. Both studies
identify self-absorption features 
between the brighter blue peak and fainter red peak of HCO$^+$ line profiles that is complemented by spectral 
peaks of optically thin lines H$^{13}$CO$^+$ and N$_2$H$^+$ line profiles at the self-absorption velocity. 
This blue asymmetric pattern is 
observed throughout the Main Ridge and supplies  
strong evidence for infalling gas along the line of sight (LOS) \citep{Myers:1995}. 
Given this evidence, we attribute the parallel alignment of the gradients and projected magnetic 
field as due to 
dominant self-gravity within the DR21 Main Ridge \citep{Lazarian:2018}. 

\subsection{Sub-filament orientations}
The sub-filaments that surround the Main Ridge of DR21 offer a different environment than the Main Ridge. While the \htwo\ column densities in the Main Ridge are mostly 
greater than 10$^{23}$~\cmsq, N(\htwo) values in the sub-filaments range 
from (5-50)$\times$10$^{21}$~\cmsq\ \citep{Pokhrel:2020}. 
The Main Ridge exhibits larger velocity dispersions with median values of 2.1~\kms\ (\co) and 
1.2~\kms\ (\coa). 
The median \co\ and \coa\ velocity dispersions in the sub-filaments 
are 1.3~\kms\ and 0.7~\kms\ respectively.   

Previous studies have proposed that the DR21 sub-filaments are 
accreting material onto the Main Ridge \citep{Schneider:2010, Hennemann:2012}.
\citet{Pillai:2026}
further demonstrated that the long axes of the sub-filaments are aligned with both the 
local gravitational acceleration direction and 
the magnetic field orientation.  Such alignments are consistent with accretion 
flows that feed both the Main Ridge and its surrounding environment.  Our analyses 
of the N and 1S filaments identify positive Rayleigh statistic values for \co.  For \coa, 
the N filament shows 
both positive and negative features in the DR21 velocity interval while 1S shows a marginal, positive 
detection.  These positive features further support the 
role of gravity in regulating the flow of gas in these regions. 

Figures 6 and 7 connect
the 3N filament with 
negative Rayleigh statistic values for both \co\ and \coa\ that indicate mostly perpendicular orientations 
between $\Psi_{\rm G,k}$ and $\Psi_{\rm B}$ within the DR21 velocity interval. 
\citet{Pillai:2026}
estimate a volume density of 1.2$\times$10$^4$~\cmsq\ for the 3N sub-filament.  
If we assume that the magnetic field strength scales with volume density,  
 as found by \citet{Crutcher:2012},
then the mean magnetic field 
strength in this sub-filament is 168~$\mu$G and the Alfv\'en velocity is 2.0~\kms. 
For a velocity dispersion of 0.94~\kms\ for this sub-filament, the Alfv\'enic 
Mach number is 0.47.  This is consistent 
with the evolution of the DR21 cloud described by
\citet{Pillai:2026}
in which the sub-filaments are sub-Alfv\'enic 
while accreting gas onto the Main Ridge. 
The sub-Alfv\'enic state of the filament enables 
Alfv\'en and slow modes to arrange 
the perpendicular orientation of  $\Psi_{\rm G,k}$ with $\Psi_{\rm B}$.

\subsection{DR21 in context with other molecular clouds}
While this paper focuses on the role of self-gravity and the interstellar magnetic field in the dynamics of the 
DR21 molecular cloud, it is useful to compare our findings with results found in 
studies of molecular clouds using similar methods.

The VChG analysis used in this study has been previously applied to the Taurus molecular cloud by both \citet{Hu:2019} and 
\citet{Heyer:2020} using 
\co\ and \coa\ J=1-0 emission and polarized dust continuum emission from Planck observations 
to determine the geometry of the interstellar magnetic field. 
\citep{Hu:2019} found mostly perpendicular alignments of the brightness temperature gradients and the magnetic field 
throughout the extent of \coa\ emission indicative of Alfv\'en 
/slow MHD modes, and sub-Alfv\'enic motions. 
\citet{Heyer:2020} found brightness temperature gradients mostly perpendicular to the local magnetic field orientation 
in the low column density envelope
of the cloud and parallel gradients in the central, high column density areas of the cloud.  This bimodal alignment follows the pattern
identified in dust column density gradients and magnetic field orientations derived from Planck data for Taurus and other clouds in the 
solar neighborhood \citep{Planck:2016}. The transition from perpendicular to parallel relative orientations
denotes the increasing role of self-gravity in the cloud dynamics \citep{Soler:2017b, Hu:2020, Seifried:2020}. 

The Orion~A molecular cloud provides a more direct comparison with DR21 as it 
is also forming young stellar clusters and massive stars.  
\citet{Zhao:2022} applied the VChG analysis using \co, \coa, and \cob\ J=1-0 data from 
the CARMA-NRO Orion Survey 
\citep{Kong:2018} and Planck observations of the 353~GHz polarized dust emission.  The derived brightness temperature gradient 
orientations are mostly perpendicular to the magnetic field orientations measured by Planck 
for each CO isotopologue.
\footnote{\citet{Zhao:2022} calculate a pseudo magnetic field orientation from the spectroscopic data by adding $\pi/2$ radians to 
the derived brightness temperature gradient orientation. We are describing their analysis in terms of the relative orientation between 
the gradient and directly measured magnetic field from the dust emission.} 
While localized sites within the Orion~A cloud are collapsing to form newborn stars and clusters such as the Orion-KL region, 
these features are washed-out by the larger number of gradient orientations that are perpendicular to the magnetic field orientations.
Accordingly, the Orion~A cloud is dominated by MHD turbulence (Alfv\'en and slow modes) throughout most of its volume. 

The relative orientations between brightness temperature 
gradients and the interstellar magnetic field found in DR21 are distinguished from 
those found in Taurus and Orion~A.
According to \citet{Lazarian:2018} and \citet{Hu:2020}, 
these differences are attributed to the dynamical states of 
the clouds.  While localized cores within the Taurus and Orion clouds are collapsing as part of 
the star formation process, both clouds 
maintain a quasi-equilibrium between self-gravity and gas pressure from 
sub-Alfv\'enic MHD turbulence that is inferred from the measured perpendicular relative 
orientations. 
The DR21 cloud is globally collapsing as found in previous studies \citep{Schneider:2010, 
Pillai:2026} and inferred from the parallel relative orientations identified in this study. 

\section{Conclusions}
In this paper we have examined the relative orientations 
between the interstellar magnetic field and spatial derivatives of antenna temperatures within thin velocity slice images of \co\ and \coa\ J=1-0 emission from the DR21 molecular cloud.
The Rayleigh statistic is computed to distinguish whether these orientations are mostly 
parallel, most perpendicular, or mostly random.  We find that the Main Ridge of the cloud shows mostly 
parallel alignment of the gradient and magnetic field position angles.  Such relative orientations are 
likely due to the large scale gravitational collapse of the cloud.  Sub-filaments N and 1S also 
show positive \PRS\ values that could be interpreted as accretion streams dominated by 
gravity.  One sub-filament, 3N, shows negative 
\PRS\ values that indicate mostly perpendicular orientations of the gradients and projected 
magnetic field and are likely a result from MHD turbulence driven by Alfv\'en and slow modes. For future 
VChG studies, we recommend using spectral line data that are mostly optically thin such as \coa\ to avoid 
line saturation that can limit \PRS\ values.

\begin{acknowledgements}
We thank the referee for their careful reading of the manuscript and analysis suggestions.
We acknowledge helpful comments from Yue Hu on the character of fast modes in the interstellar medium. 
This work made use of Astropy, a community-developed core Python package and an ecosystem of tools and resources for astronomy \citep{astropy:2013, astropy:2018, astropy:2022}.  
This paper is based on observations made with the NASA/DLR Stratospheric Observatory for Infrared Astronomy (SOFIA).  SOFIA was jointly operated by the 
Universities Space Research Association, Inc. (USRA), under NASA contract
NNA17BF53C and the Deutsches SOFIA Institut (DSI) under DLR contract 50~OK~0901 to the University of Stuutgart.

MH acknowledges financial 
support from MIT grant S5974.  TGSP acknowledges support by NASA award \#09-0215
issued by USRA and awards from the National Science Foundation under grant AST-2009842 and AST-2108989.
DS acknowledges support from Bonn-Cologne Graduate School, which is funded through the German Excellence initiative as well as funding by the Deutsche 
Forshungsgemeinschaft (DFG) via the Collaborative Research Center SFB 1601
``Habitats of massive stars across cosmic time" (subprojects B1 and B4).
This work was carried out in part at the Jet Propulsion Laboratory, 
which is operated by the California Institute of Technology under a 
contract with the National Aeronautics and Space Administration (80NM0018D0004).
\end{acknowledgements}

\bibliographystyle{aa}
\bibliography{dr21_hro.bib}

\begin{appendix}
\section{Impact of line opacity on the measured gradients}
\begin{figure}[htb]
\begin{center}
\includegraphics[width=0.4\textwidth,trim={0 0 0 0}, clip]{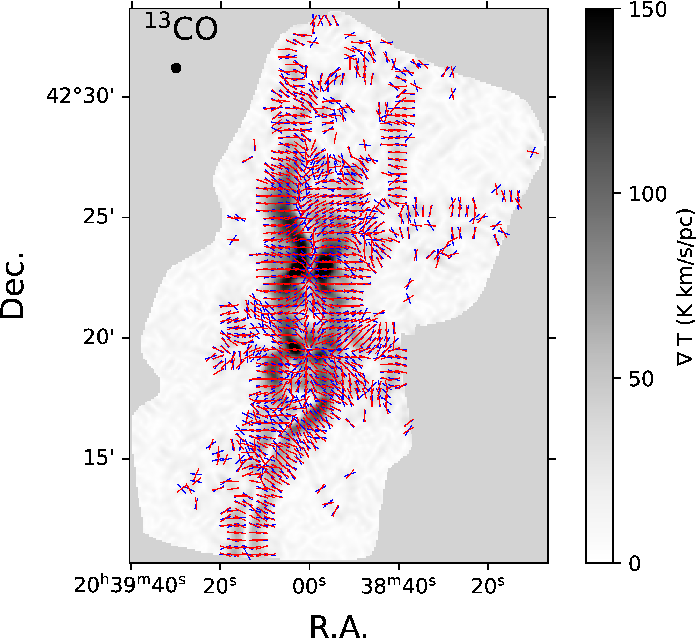}
\caption{Image of gradients derived from \coa\ integrated intensities.  The blue and red segments show 
gradient orientations derived from \coa\ integrated intensities (blue) and \htwo\ column densities (red).  Only segments 
from pixels with gradient magnitude signal to noise ratios greater than 3.5 are shown.  The filled, black 
circle represents the full width at half maximum beam size. 
}
\label{A1}
\end{center}
\end{figure}

In this section, we present the gradient orientations derived from \coa\ integrated intensities and \htwo\ column densities that are discussed 
in Sect.~4. 
DR21 \htwo\ column densities are calculated from modified Black body fits to multi-wavelength dust emission observed with {\it Herschel} 
over the Cygnus~X complex \citep{Pokhrel:2020}. 
It is assumed that the dust emission is optically thin throughout the DR21 cloud area. 
\coa\ integrated intensites are calculated over the velocity interval [-7,1.5]~\kms. 
Figure~\ref{A1} shows the gradients derived from \coa\ integrated intensities and the 
blue and red segments show gradient orientations for \coa\ and \htwo\ column density respectively.  
Unlike the \coa\ velocity channel 
images in Figure~\ref{figure2} and Figure~\ref{figure5}, most areas within the main cloud satisfy the 
gradient magnitude signal to noise 
ratio of 3.5.   
The gradient orientations 
derived from each tracer
are well aligned in parallel with each other.  This is confirmed by the derived \PRS\ value of 63.  
While the optical depth of \coa\ emission may exceed values greater than 1 in some dense, localized areas, there is no evidence of 
significant line saturation over most of the area of the cloud as found in the \co\ data. 

\end{appendix}

\end{document}